\documentclass[a4paper]{article}

\usepackage[margin=25mm]{geometry}
\usepackage{amsmath}
\usepackage{amsfonts}
\usepackage{amssymb}
\usepackage{graphicx}
\usepackage[utf8]{inputenc}
\usepackage{authblk}
\usepackage{imakeidx}
\usepackage{tabularx}
\usepackage[sorting=none]{biblatex}

\DeclareUnicodeCharacter{0301}{\'{e}}
\DeclareUnicodeCharacter{0413}{\'{e}}
\DeclareUnicodeCharacter{043E}{\'{e}}
\DeclareUnicodeCharacter{043B}{\'{e}}
\DeclareUnicodeCharacter{043F}{\'{e}}
\DeclareUnicodeCharacter{0440}{\'{e}}
\DeclareUnicodeCharacter{0438}{\'{e}}
\DeclareUnicodeCharacter{0441}{\'{e}}
\DeclareUnicodeCharacter{0442}{\'{e}}
\DeclareUnicodeCharacter{0430}{\'{e}}
\DeclareUnicodeCharacter{043D}{\'{e}}
\DeclareUnicodeCharacter{044C}{\'{e}}
\DeclareUnicodeCharacter{043A}{\'{e}}
\DeclareUnicodeCharacter{0439}{\'{e}}
\DeclareUnicodeCharacter{043C}{\'{e}}
\DeclareUnicodeCharacter{0456}{\'{e}}
\DeclareUnicodeCharacter{0446}{\'{e}}
\DeclareUnicodeCharacter{0435}{\'{e}}
\DeclareUnicodeCharacter{0437}{\'{e}}
\DeclareUnicodeCharacter{044F}{\'{e}}

\providecommand{\keywords}[1]
{
  \small	
  \textbf{\textit{Keywords---}} #1
}

\title{MAPD type avalanche photodetectors}

\author[1]{K. Isayev}
\author[2]{A. Mammadli}
\author[2]{K. Khuseynzada}
\author[2]{A.Sadigov}
\affil[1]{National Nuclear University MEPHI, Moscow, Russia}
\affil[2]{National Nuclear Research Center, Baku, Azerbaijan}

\begin{document}

\maketitle

\begin{abstract}
Recent years getting progress in the area of micropixel avalanche photodiode makes them attractive as sensors for scintillation detectors. The performance of the scintillation detectors is strongly tied to the physical properties of MAPD. For this point of wives, we will investigate three types MAPD-3NK, MAPD-3NM and MAPD-3NM-II  for possible use in a scintillation detector. Investigated avalanche photodiodes have the following parameters e.g low operation voltage (90V$<$U$<$50V), high pixel density (4000 pixel /$mm^2$$<$Pixel density$<$10000 pixel/$mm^2$), high photodetection efficiency (20\% $<$ PDE $<$40\%) and high gain (6*$10^4$$<$M$<$3*$10^5$). Gamma-ray detection performance of tested scintillation detector based on MAPD was checked with different gamma rays. Obtained results to confirm that this kind of scintillation detectors can be successfully used for the next generation of PET cameras in medicine.
\end{abstract}

\keywords{SiPM; MAPD; silicon photomultiplier; avalanche photodiode; buried pixel.}

\section{Introduction}
Rapid progress in the physics and technology of semiconductors in the last century led to the creation of solid-state analogs of almost all vacuum devices, except for photomultiplier tubes (PMTs). Although, after the discovery in the early 50s of the last century of the effect of increasing the photocurrent in silicon and germanium avalanche photodiodes and the creation of the basic theory of impact ionization in semiconductors, certain prospects appeared, but the creation of adequate solid-state analogs of PMTs required many years of research. The avalanche photodiodes (APDs) known at that time were significantly inferior to PMTs in such basic parameters as the gain, the size of the working area, and the threshold sensitivity. In the 90s, Silicon Photoelectron Multipliers (SiPM) detectors were developed \cite{sadygov2020silicon}, \cite{sadygov2018photodetector}, \cite{sadygov2014possibilities}, \cite{sadygov2014microchannel} and are now widely used, capable of registering single photons at room temperature. SiPM detectors operate in a mode above the breakdown voltage (or in the Geiger mode) and have a wide range of photoresponse linearity in terms of the number of photons per pulse \cite{ahmadov2017new}, \cite{ahmadov2022investigation}. Depending on the manufacturer, SiPMs are also called Micro Pixel Avalanche Photodiodes (MAPD) or Micro Pixel Photon Counters (MPPC) \cite{sadygov2016multi}. At present, SiPM detectors in their main parameters, such as the efficiency of photon detection, gain, region of linearity of the photoresponse, and working area, significantly exceed SPAD and VLPC detectors. Therefore, the SiPM detector is considered the most adequate solid-state analog of the PMT. During the past decade, mass production of SiPM detectors has been achieved. So far, the main consumers of SiPM detectors are large-scale projects in high-energy physics carried out in leading scientific centers of the world, including the NICA project (Dubna, Russia) \cite{holik2020program}, \cite{holik2020miniaturized}, \cite{akbarov2020scintillation}. Tests are being carried out with the aim of mass application of SiPM detectors in medical tomographs of a new generation and in the automotive industry \cite{nazarov2017design}. You can notice the fact that the structures and designs of SiPM are constantly being modernized, thereby improving their operating parameters \cite{nuriyev2018performance}, \cite{sadigov2022improvement}. It is required to carry out research on the parameters of new silicon photodiodes, and also to conduct an analytical study with other analogs \cite{ahmadov2014micro}, \cite{ahmadov2014new}, \cite{holik2022gamma}.

\section{Structure}

The design of the structure of an avalanche photodiode with a buried arrangement of pixels is shown in Fig. 1. It is a silicon substrate with n-type conductivity, on the surface of which two epitaxial layers of p-type conductivity are grown. The resistivity of epitaxial layers is usually chosen in the range 5–10 Ohm * cm. The thickness of the epitaxial layers is chosen by 4 um each. Between the epitaxial layers, a matrix of highly doped n plus -type regions is formed with a step of 5 to 15 um, depending on the specific design. This provides an increase in the pixel density up to 40,000 pixels / $mm^2$, with a 100\% sensitive area of the device. The manufacturing technology of the MAPD is described in \cite{sadigov2017micropixel}, \cite{jafarova2014features}, \cite{khuseynzada2022innovative}.

\begin{figure}[ht]
   \centering
   \includegraphics[width=0.6\textwidth]{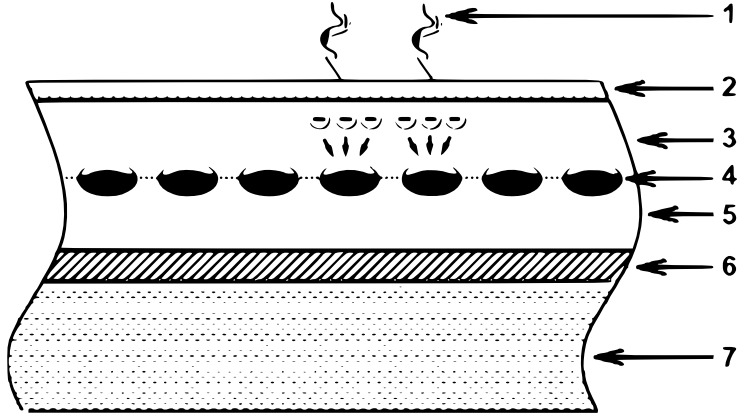} 
   \caption{Cross-section of avalanche photodiodes with deeply hidden pixels: 1– registered photons, 2 - highly doped p plus - layer to provide contact to the epitaxial layer, 3 - second epitaxial p-type layer, 4 - potential wells (pixels) from n plus -regions, 5 -first p-type epitaxial layer, 6 - highly doped n plus -layer to limit the depletion region, 7 - n-type silicon substrate.}
    \label{fig1}
\end{figure}

One of the advantages of this that in this structure it can achieve a much higher pixel density than in the SiPMs with surface pixels arrangement. This is due to the fact that the buried pixel structure does not have a common conducting bus and individual resistors, which could occupy a significant part of the sensitive area with an increase in the pixel density. The function of a quenching individual resistor is performed by a forward-biased p-n-junction located under each pixel \cite{yilmazelectrical}, \cite{sadigovaimprovement}.

\section{Working principle}
In the operating mode, a negative voltage is applied to the SiPM relative to the substrate. Depletion of the device begins with the first p-n junction located at the between the substrate and the first epitaxial layer. At a certain voltage value, the depletion region reaches the matrix of n plus -regions and partially opens the second p-n-junction available there. From this moment on, only the third p-n-junction, located at the border of n plus -regions with the second epitaxial layer, begins to deplete. A further increase in voltage leads to a complete depletion of the second epitaxial layer. As a result, a matrix of potential wells of n plus -regions is formed in the depletion region of the SiPM, and a hemispherical electric field is formed above each of these regions, which ensures the collection of photoelectrons from the entire sensitive surface of the device. Thus, the sensitive surface of the device is divided into photosensitive regions with individual pixels (or multiplication micro channels), completely independent of each other.

The avalanche multiplication of charge carriers in microchannels occurs in the near-boundary region of the second epitaxial layer with n plus -regions, where high electric field strength is generated. The multiplied electrons accumulate in potential wells formed by n plus -regions, which leads to a decrease in the electric field in the second epitaxial layer below a certain threshold value, as a result of which the avalanche process in this channel stops.
The restoration of the previous field in the multiplication microchannel occurs due to the accumulated charge draining into the volume of the substrate through the directly biased p-n junction formed between the first epitaxial layer and the n plus -region.

\section{Experiment and results}

The 7 samples with different designs of deeply submerged pixel structures of the MAPD were compared. All samples were tested under the same laboratory conditions. The basic working parameters of the photodiode performance have been determined. An experimental setup for determining the breakdown and operating voltages of samples is schematically shown in Figure 2. 

\begin{figure}[ht]
    \centering
    \includegraphics[width=0.4\textwidth]{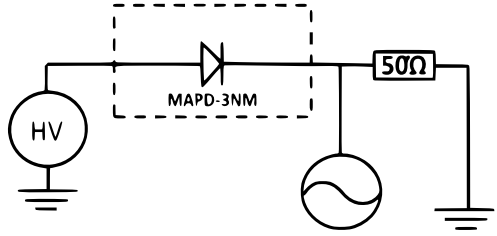}
    \caption{Experimental setup for breakdown and operation voltage determination.}
    \label{fig2}
\end{figure}

The sample was connected to a power source, which was a KEITHLEY 6487 picoammeter. The entire setup was isolated from the effects of extraneous light. The picoammeter was used to find the values of the dark current $I_d$ for a given voltage.

\begin{figure}[ht]
    \centering
    \includegraphics[width=0.9\textwidth]{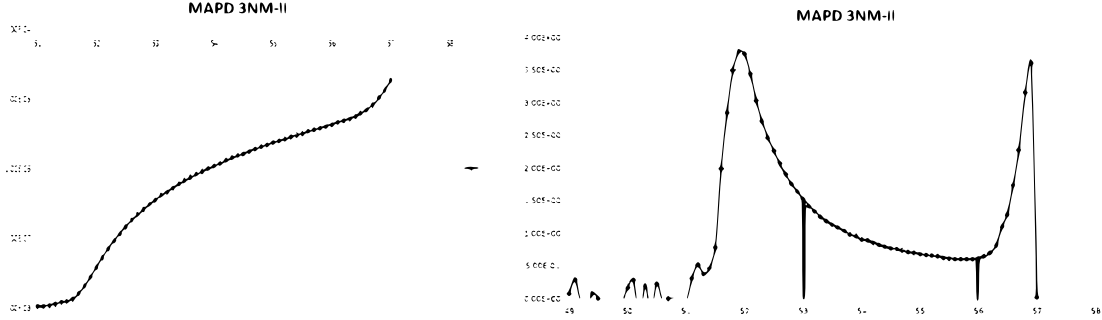}
    \caption{V-A characteristic of the MAPD-3NM-I sample.}
    \label{fig3}
\end{figure}

It was determined that the breakdown voltages $U_br$ were in the range 51.8 - 52 V, and the operating voltage $U_op$ was 53 - 56V (Figure 3). The measurements were carried out at a temperature of T was 21 C.

One of the main parameters of photodiodes is the photon detection efficiency (PDE). The sample was illuminated by a light diode with a wavelength of 450 nm, to which a 30 ns pulse was applied from a generator with a frequency of 50 kHz and amplitude of 2.92 V. The measurements were carried out at a temperature of \text{-}6 °C, taking into account the change in breakdown voltage of samples.

Their comparative analysis was carried out. The obtained results are shown in table below.

\begin{tabularx}{0.8\textwidth} { 
  | >{\raggedright\arraybackslash}X 
  | >{\centering\arraybackslash}X 
  | >{\centering\arraybackslash}X 
  | >{\centering\arraybackslash}X | }
  
 \hline
MAPD & 3NK & NM-I & NM-II \\
 \hline
Production year  & 2017  & 2020  & 2021 \\
  \hline
Pixel density  & 10000  & 10000 & 10000 \\
   \hline
Size mm  & 3.7*3.7  & 3.7*3.7 & 3.7*3.7  \\
\hline
Photon detection efficiency (450-550) \%  & 32 & 33 & 36 \\
\hline
Operation voltage V  & 91 & 74 & 55 \\
\hline
Gain *$10^4$  & 6 & 12 & 16  \\
\hline
Dark current nA  & 1000 & 220 & 100  \\
\hline
\end{tabularx}

\section{Conclusion }
As a result of the research done, the structures and principles of operation of silicon photodiodes were studied. The main parameters characterizing the SiPM were determined. As part of this work, breakdown and operating voltages, dark current, gain and PDE were determined for the MAPD-3NM. The breakdown voltage $U_br$ was found to be 51.8 - 52V, the operating voltage $U_op$ has values in the range of 53 - 56 V.  The gain for the MAPD-3NM-II was 1.6 * $10^5$.

\section*{Acknowledgement}
This work was supported by the Scientific Foundation of SOCAR. 

\printbibliography
\end{document}